\documentclass[final,5p,times,twocolumn]{elsarticle}

\usepackage{graphics}
\usepackage{amssymb}
\usepackage{textcomp}
\begin{document}

\begin{frontmatter}



\title{Pseudogap Phase Boundary in Overdoped Bi$_{2}$Sr$_{2}$CaCu$_{2}$O$_{8+\delta}$ Studied by Measuring Out-of-plane Resistivity under the Magnetic Fields}


\author{\underline{Kousuke Murata$^{1}$}, Haruki Kushibiki$^{1}$, Takao Watanabe$^{1}$,
Kazutaka Kudo$^{2}$, Terukazu Nishizaki$^{2}$, Norio Kobayashi$^{2}$, \\ Kazuyoshi Yamada$^{3}$, Takashi Noji$^{4}$, and Yoji Koike$^{4}$}

\address{$^{1}$Graduate School of Science and Technology, Hirosaki University, 3 Bunkyo, Hirosaki, 036-8561 Japan

$^{2}$Institute for Materials Research, Tohoku University, 2-1-1 Katahira, Aoba-ku, Sendai, 980-8577 Japan

$^{3}$Advanced Institute for Materials Research (WPI), Tohoku University, 2-1-1 Katahira, Aoba-ku, Sendai, 980-8577 Japan

$^{4}$Graduate School of Engineering, Tohoku University, 6-6-05 Aoba, Aramaki, Aoba-ku, Sendai, 980-8579 Japan

Email: twatana@cc.hirosaki-u.ac.jp}

\begin{abstract}
In order to elucidate the relationship between the pseudogap and high-$T_{c}$ superconductivity, the characteristic pseudogap temperature $T^{*}$ in Bi$_{2}$Sr$_{2}$CaCu$_{2}$O$_{8+\delta}$ system has been systematically evaluated as a function of doping, especially focusing on its overdoped region, by measuring the out-of-plane resistivity $\rho_{c}$. Overdoped samples have been prepared by annealing TSFZ-grown Bi$_{2}$Sr$_{2}$CaCu$_{2}$O$_{8+\delta}$ single crystals under the high oxygen pressures (990 kgf/cm$^{2}$). At a zero field, the $\rho_{c}$ showed a metallic behavior down to $T_{c}$(= 62 K), while under the magnetic fields of over 3 T, $\rho_{c}$ showed typical upturn behavior from around 65 K upon decreasing temperature. This result suggests that the pseudogap and superconductivity are different phenomena.
%%This implies that the pseudogap temperature $T^{*}$ (= 65 K) which has been hidden by the onset of superconductivity at a zero field has emerged along with the suppression of superconductivity by the application of high magnetic fields. 

\end{abstract}


\end{frontmatter}
%%\linenumbers


\section{Introduction}
It is well recognized that an understanding of the pseudogap phenomena in high-$T_{c}$ cuprates is crucially important to understand the mechanism of high-$T_{c}$ superconductivity. It has not been yet clarified, however, whether it is a precursor of high-$T_{c}$ superconductivity or it is some kind of a competing order against the superconductivity. In the former case, the pseudogap temperature $T^{*}$ may decrease with increasing doping and merge with $T_{c}$ in the overdope state. In the latter case, the $T^{*}$ may cross the $T_{c}$ dome and point to zero at the quantum critical point. Thus, in order to clarify the origin of the pseudogap, it is important to investigate the $T^{*}$ as a function of doping. Experimental data in the heavily overdoped state is lacking, however, especially on Bi$_{2}$Sr$_{2}$CaCu$_{2}$O$_{8+\delta}$ (Bi-2212) due to the difficulty in sample preparations. 
%%A phenomenon that reduces the density of state at low energy called the pseudogap is observed generally in the high-$T_{c}$ superconductivity, it is well recognized that an understanding of pseudogap closely related  with superconducting mechanism. There is the greatest issue whether a pseudogap means the formation of the electron pair of the superconduction origin, or whether it means some kind of order formation to compete with the superconduction. In the first case, it is thought that pseudogap temperature $T^{*}$ decrease with doping of the carrier and accords $T_{c}$ in the overdope state. In the latter case, $T^{*}$ crosses $T_{c}$, and it is expected that $T^{*}$ go to the absolute zero point (ground state) called a quantum critical point. If we can make clear the electronic phase diagram by the experiment, lead to an understanding of the mechanism of high-$T_{c}$ temperature superconductivity and pseudogap.

Among various techniques to detect the pseudogap effect, the out-of-plane resistivity $\rho_{c}$ is one of the most powerful method, since it directly probes the electronic density of states (DOS) around the Fermi level. Then, the $\rho_{c}$ shows a typical upturn below the pseudogap temperature $T^{*}$~\cite{watanabe}.  Recently, the $\rho_{c}$ of Bi-2212 has also been measured under the high magnetic fields up to 60 T and a negative magnetoresistance has been observed below the $T^{*}$~\cite{t.shiba}. This result has been attributed to the suppression of the pseudogap effect under the high magnetic fields.

In this paper, we extended the doping level of Bi-2212 to the heavily overdoped state and measured the out-of-plane resistivity $\rho_{c}$ under the various magnetic fields. Based on the results, we discuss the pseudogap phase boundary in the overdoped Bi-2212. 
%%Recently, the pseudogap has been actively researched, for example, it is reported that the increase of out-of-plane resistivity $\rho_{c}$ happene with pseudogap by resistivity measurements under magnetic field up to 60T in Bi$_{2}$Sr$_{2}$CaCu$_{2}$O$_{8+\delta}$ (Bi-2212) crystal [1]. This phenomenon happens sensitive to the pseudogap, so the experiment of the c-axis resistivity measurements i s a good method in the research of the pseudogap. On the other hand, it is known that Bi-2212 single crystals is changed a carrier by amount of oxygen control, but the data of overdopedope area that $T^{*}$ nears $T_{c}$ have few report because overdope samples preparing is difficult. In this research, we intended to determine the origin of the pseudogap, we report that we investigated behavior of $T^{*}$ from the experiment of the c-axis resistivity measurements under the magnetic field.the $\rho_{c}$The pseudogap formation cause a typical upturn of the $\rho_{c}$  below the $T^{*}$.

\label{}


\section{Experimental}
Single crystals were grown using the traveling solvent floating zone method~\cite{t.watanabe}. Heavily overdoped samples with $T_{c}$ (= 62 K) were made by high O$_{2}$ pressure (990 kgf/cm$^{2}$) annealing at 500 \textcelsius~  for 100 h. The c-axis length was estimated from the X-ray diffraction pattern using fitting method of Nelson-Riley function. The c-axis length of the annealed sample shrinked about 0.15 \AA ~compared with that of the as-grown sample, confirming that excess oxygen was actually incorporated into the sample~\cite{t.fujii}. The doping level $p$ was estimated as 0.22 using the empirical relation proposed by J. L. Tallon~\cite{tallon}. $\rho_{c}$ measurements were carried out by a DC four-terminal method. Magnetic fields were applied up to 17.5 T parallel to the c-axis by a superconducting magnet.  
%%We used Bi-2212 single crystal as samples. Bi-2212 have been prepared by TSFZ-grown. And overdoped samples have been prepared by annealing under the high oxygen pressures. In order to change the dope quantity of the samples, we annealed have been prepared under various pressures. Grown crystals were characterized using X-ray diffraction (XRD). In order to estimate the c-axis length, we used a fitting method with Nelson-Riley (N-R) function. c-axis resistivity under the magnetic field used direct current four-terminal method. We impressed the magnetic field to 17.5 T  and measured $\rho_{c}$. Impression angle of the magnetic field set H//c. The formula of J. L. Tallon was used to decision of doping quantity p of the sample [2].

\begin{figure}[t] 
\includegraphics[width=9cm,clip]{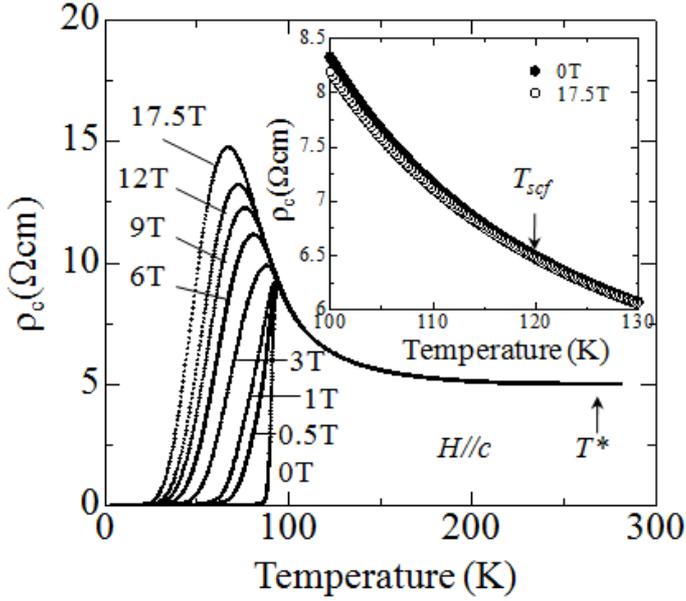}
\caption{Out-of-plane resistivity $\rho_{c}$, of an Bi$_{2}$Sr$_{2}$CaCu$_{2}$O$_{8+\delta}$ single crystal under the various magnetic fields. The inset shows the expanded plots of the out-of-plane resistivity $\rho_{c}$ just above $T_{c}$.}

\end{figure}

\section{Results and Discussion}
Figure 1 shows the temperature dependence of out-of-plane resistivity $\rho_{c}$ for an optimally doped Bi-2212 single crystal under various magnetic fields. At a zero field, the $\rho_{c}$ is semiconductive in all temperature regions measured. The $T^{*}$ can safely be estimated as around room temperature. When we applied magnetic fields, we observed negative magnetoresistance at low temperatures near $T_{c}$. The inset of Fig. 1 shows the enlarged $\rho_{c}$ just above $T_{c}$. The data at the magnetic field of 17.5 T deviates downward from that at a zero field below around 120 K, indicating the occurrence of the negative magnetoresistance. 

Although the typical upturn of the $\rho_{c}$ is generally thought to be caused by the opening of the pseudogap~\cite{watanabe}, this behavior can also be caused by the superconductive fluctuation effect~\cite{l.b.ioffe}, if the material has an electronically extreme 2D nature such as high-$T_{c}$ cuprates. This is because the in-plane DOS reduction by the superconductive fluctuation affects the $\rho_{c}$ in a similar manner as that by the pseudogap. Then, when the superconductivity is suppressed by the magnetic fields, the DOS is expected to be recovered and cause the negative magnetoresistance. Consequently, the estimated temperature, 120 K, for the negative magnetoresistance may be the onset temperature, $T_{scf}$, for the superconductive fluctuation. Indeed, the temperature coincides with that estimated by the measurements of static susceptibilities~\cite{watanabe}. The observed sharp increase in $\rho_{c}$ near $T_{c}$ may be attributed to both the pseudogap and the superconductive fluctuation effects.

Figure 2 shows the temperature dependence of out-of-plane resisitivity $\rho_{c}$ for the annealed sample under various magnetic fields. At a zero field, the $\rho_{c}$ showed a metallic behavior down to $T_{c}$(= 62 K), indicating the sample is in a heavily overdoped state, while under the magnetic fields of over 3 T, the $\rho_{c}$ showed typical upturn behavior from around 65 K upon decreasing temperature. This implies that the pseudogap temperature $T^{*}$ (= 65 K) which has been hidden by onset of superconductivity at a zero field has emerged along with the suppression of superconductivity by the application of the high magnetic fields. On the other hand, $T_{scf}$ could not be found for this sample. If the $T_{scf}$ scales with $T_{c}$, the $T_{scf}$ may be higher than observed $T^{*}$. This is probably because the electronic 2D nature, which is necessary to observe the upturn of $\rho_{c}$ due to the superconductive fluctuation effect, become weak in the heavily overdoped state. This result suggests that the  $T^{*}$ crosses the $T_{c}$ dome in the heavily overdoped Bi-2212.

\begin{figure}[t]
\includegraphics[width=9cm,clip]{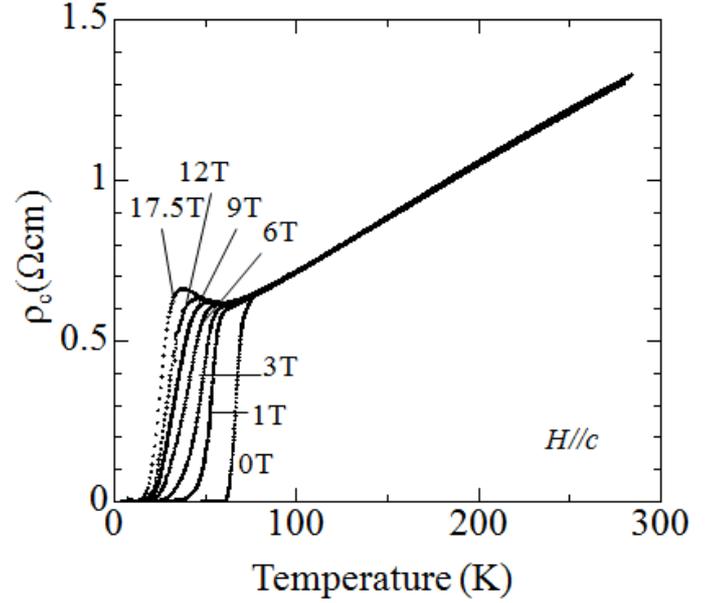}
\caption{Out-of-plane resistivity $\rho_{c}$, of an heavily overdoped Bi$_{2}$Sr$_{2}$CaCu$_{2}$O$_{8+\delta}$ single crystal under the various magnetic fields.}
\end{figure}

\section{Conclusion}
In order to understand the pseudogap phenomena more in detail, the pseudogap temperatures $T^{*}$ of the Bi-2212 single crystals have been investigated by the measurements of the out-of-plane resistivity $\rho_{c}$ under the magnetic field of up to 17.5 T. The results of an optimally doped sample indicate that, although the typical upturn in $\rho_{c}$ first occurs at $T^{*}$ due to the opening of the pseudogap, the superconductive fluctuation effect is added to the behavior at temperatures near $T_{c}$.  Moreover, the results of the heavily overdoped sample indicate that the  $T^{*}$ seem to cross the $T_{c}$ dome. Both these results suggest that the pseudogap and superconductivity are different phenomena.

\end{document}